\documentclass[%
 reprint,
 superscriptaddress,
nofootinbib,
 amsmath,amssymb,
 aps,
]{revtex4-2}
\usepackage{graphicx}
\usepackage{dcolumn}
\usepackage{bm}
\usepackage{color}
\usepackage{gensymb}
\usepackage{hyperref}
\usepackage[normalem]{ulem}

\usepackage{aas_macros}

\begin{document}

\title{Posterior predictive checking for gravitational-wave detection with pulsar timing arrays: I. The optimal statistic}

\author{Michele Vallisneri}
\email{michele.vallisneri@jpl.nasa.gov}
\affiliation{Jet Propulsion Laboratory, California Institute of Technology, Pasadena CA 91109, USA}
\affiliation{Department of Physics, California Institute of Technology, Pasadena, California 91125, USA}

\author{Patrick M. Meyers}
\email{pmeyers@caltech.edu}
\affiliation{Department of Physics, California Institute of Technology, Pasadena, California 91125, USA}
\author{Katerina Chatziioannou}
\email{kchatziioannou@caltech.edu}
\affiliation{LIGO  Laboratory,  California  Institute  of  Technology,  Pasadena,  California  91125,  USA}
\affiliation{Department of Physics, California Institute of Technology, Pasadena, California 91125, USA}
\author{Alvin J. K. Chua}
\email{alvincjk@nus.edu.sg}
\affiliation{Department of Physics, National University of Singapore, Singapore 117551}
\affiliation{Department of Mathematics, National University of Singapore, Singapore 119076}

\date{\today}

\begin{abstract}
A gravitational-wave background can be detected in pulsar-timing-array data as Hellings--Downs correlations among the timing residuals measured for different pulsars.
The \emph{optimal statistic} implements this concept as a classical null-hypothesis statistical test: a null model with no correlations can be rejected if the observed value of the statistic is very unlikely under that model.
To address the dependence of the statistic on the uncertain pulsar noise parameters, the pulsar-timing-array community has adopted a hybrid classical--Bayesian scheme (Vigeland et al.\ 2018) in which the posterior distribution of the noise parameters induces a posterior distribution for the statistic.
In this article we propose a rigorous interpretation of the hybrid scheme as an instance of posterior predictive checking, and we introduce a new summary statistic (the Bayesian signal-to-noise ratio) that should be used to accurately quantify the statistical significance of an observation instead of the mean posterior signal-to-noise ratio, which does not support such a direct interpretation. 
In addition to falsifying the no-correlation hypothesis, the Bayesian signal-to-noise ratio can also provide evidence supporting the presence of Hellings--Downs correlations.
We demonstrate our proposal with simulated datasets based on NANOGrav's 12.5-yr data release.
We also establish a relation between the posterior distribution of the statistic and the Bayes factor in favor of correlations, thus calibrating the Bayes factor in terms of hypothesis-testing significance.
\end{abstract}

\maketitle


\section{Introduction and summary}

 In June 2023, four international pulsar-timing-array (PTA) efforts (NANOGrav, EPTA, PPTA, and CPTA) reported compelling evidence \cite{NANOGrav:2023gor,EPTA:2023fyk,Reardon:2023gzh,Xu:2023wog} for the existence of a low-frequency gravitational-wave background (GWB), as expected from the binaries of supermassive black holes at galactic centers \cite{NANOGrav:2023hfp,EPTA:2023xxk,EPTA:2023gyr,NANOGrav:2023pdq}, although more exotic sources are also possible \cite{NANOGrav:2023hvm,EPTA:2023xxk}.
These results had been prefigured by findings of excess timing noise \cite{abb+20,ppta_dr2_gwb,epta_dr2_gwb,2022MNRAS.510.4873A} with spectra that were consistent across pulsars and plausible with respect to astronomical expectations~\cite{2021MNRAS.502L..99M}.
These findings were suggestive but by no means conclusive, since the excess noise may have arisen from sources other than the GWB, such as intrinsic pulsar spin noise~\cite{zhs+22,gts+22}.

Indeed, in Refs.\ \cite{NANOGrav:2023gor,EPTA:2023fyk,Reardon:2023gzh,Xu:2023wog} evidence for the GWB was established using a different criterion: the presence of specific correlations between the time series of residuals of different pulsars.
For an isotropic GWB, these correlations follow a geometric relation (a function of the angle $\zeta$ between the pulsars) first derived by Hellings and Downs~\cite{hd83}. 
While other systematic effects (such as clock and ephemeris errors~\cite{thk+2016,2020ApJ...893..112V}) may generate different angular patterns, it is difficult to imagine an explanation other than the GWB for a manifest Hellings--Downs pattern in the data.

The problem of GWB detection with PTAs data then turns into the probabilistic characterization of observed inter-pulsar correlations.
Bayesian approaches have been the tool of choice, since they allow for a principled treatment of all unknown variables needed to fully describe the data, such as the geometric and kinematic parameters of the pulsars, the levels and spectral shapes of radiometric noise, intrinsic spin noise, and dispersion-measure noise from the interstellar medium, and of course the GWB parameters (see Refs.\ \cite{2009MNRAS.395.1005V,taylor2021nanohertz} and references therein). 
In the dominant Bayesian approach, the statistical evidence in favor of Hellings--Downs correlations is quantified as the Bayes factor between two data models that include all the elements listed above, and that are identical \emph{except} for the inter-pulsar correlation coefficients of the shared common-spectrum process, which are set to zero for the null model (common-spectrum uncorrelated red noise, or \textsc{curn}) and to the Hellings--Downs pattern for the isotropic-GWB model (\textsc{hd}).
Thus, this model comparison begins with the well-established finding of excess timing-residual power, and attempts to attribute its origin to either independent processes in each pulsar, or to the phase-coherent delays induced by the GWB.
See Refs.~\cite{2018ApJ...859...47A,abb+20} for examples of this analysis.
A complementary Bayesian strategy based on posterior predictive model checking \cite{gelman2013bayesian} is discussed in a companion to this paper \cite{ppc2}.

By contrast, classical frequentist statistics offers an attractive formulation of GWB detection based on \emph{null hypothesis statistical testing} \cite{pernet2015}, as used in LIGO's historical black-hole binary detection \cite{2016PhRvL.116f1102A}.
For GWB searches with PTAs, null hypothesis testing is implemented with the \emph{optimal statistic} \cite{2009PhRvD..79h4030A,2015PhRvD..91d4048C,2018PhRvD..98d4003V}, which was devised as a direct measure of Hellings--Downs correlations.
The basic idea is that if we observe a value of the optimal statistic much larger than expected under the null hypothesis of no inter-pulsar correlations, we can reject that scenario and conclude that correlations are present.
The $p$-value (the probability of obtaining the observed value of the optimal statistic, or larger, under the null hypothesis) quantifies the statistical significance of this conclusion.

Unfortunately, this simple test cannot be implemented in practice because the optimal statistic depends parametrically on the unknown pulsar noise parameters.
\citet{2018PhRvD..98d4003V} introduced a hybrid scheme in which the posterior distribution of pulsar noise parameters (obtained from Bayesian inference) induces a distribution of the observed optimal statistic, and proposed that the posterior mean of the signal-to-noise ratio (SNR)---i.e., of the optimal statistic divided by its standard deviation across noise realizations---could be used as a measure of statistical significance. However, the mean posterior SNR does not correspond to a $p$-value for the optimal statistic, so its hypothesis-testing interpretation is questionable. Partly because of this, and because of the perception that Bayesian model comparison accounts more fully for the uncertainties in the data model, the optimal statistic has been relegated to a secondary role in PTA searches for the GWB.

In this article we submit that the optimal statistic should remain a tool of first choice in these searches, on par with Bayesian model comparison.
Specifically, in Sec.~\ref{sec:formalism} we show that the hybrid optimal statistic can be interpreted rigorously in the framework of posterior predictive model checking~\cite{gelman2013bayesian}, leading to a self-consistent Bayesian generalization of statistical testing that can falsify the no-correlation hypothesis while accounting for the uncertainty in pulsar noise parameters. We introduce a new detection statistic (the \emph{Bayesian SNR}, henceforth BSNR) that maps to a single well-defined $p$-value and provides a direct measure of statistical significance.
In Sec.~\ref{sec:simulation} we demonstrate this scheme by examining the distribution of hybrid SNRs and BSNR in two sets of simulated datasets (with and without a loud GWB injection).
In Sec.~\ref{sec:alternative} we discuss the role that the BSNR can play in discriminating a true Hellings--Downs-correlated signal from spurious processes with other spatial correlations.
In Sec.~\ref{sec:plr} we exploit the fact that the (squared) optimal statistic approximates the log ratio of the \textsc{hd} and \textsc{curn} likelihoods to formulate detection statements based on the distribution of the posterior log likelihood ratio (PLLR, \cite{dempster1997direct,2011AIPC.1305..391S}). We show that the PLLR is related to the \textsc{hd}-vs.-\textsc{curn} Bayes factor, providing a frequentist calibration for its value, and confirming that the commonly used $\exp(\mathrm{SNR}^2/2)$ heuristic for the Bayes factor can overpredict its value.
In Sec.~\ref{sec:conclusion} we present our conclusions.

\section{Null hypothesis tests and the Bayesian $p$-value}
\label{sec:formalism}

Classical null hypothesis testing \cite{pernet2015} proceeds by assigning a function $D$ of the data $y$ to serve as a \emph{test statistic}, and then rejecting the null hypothesis $H_0$ when the observed value $D_\mathrm{obs} = D(y_\mathrm{obs})$ is in the extreme tail of its background distribution 
\begin{equation}
p(D|H_0) = \int p(D|y) p(y|H_0) \, \mathrm{d}y\,;
\end{equation}
that is, when it is very unlikely that $H_0$ could produce data that result in $D_\mathrm{obs}$.
The tail area $P(D > D_\mathrm{obs}|H_0)$ (or $P(D < D_\mathrm{obs}|H_0)$ as appropriate) is known as the one-sided $p$-value.
Just how small a $p$-value should justify the rejection of the null hypothesis depends on extra-statistical considerations, and has been the subject of considerable debate.
Crucially, the $p$-value quantifies the probability that $H_0$ would generate the observed data, and \emph{not} the probability that $H_0$ is true given the data, which depends on the \emph{base rate} of $H_0$ in ``similar'' experiments. 

In the PTA context, the optimal statistic is used as follows to implement null hypothesis testing:
\begin{itemize}
\item we construct the optimal statistic (which we will again denote as $D$) specifically to quantify the strength of Hellings--Downs correlations in the data; by design, the optimal statistic embodies a fiducial GWB spectral shape (but not its overall amplitude), as well as a fiducial noise model for each pulsar;
\item we compute the background distribution $p(D|H_0)$ under the null hypothesis $H_0$ ($\equiv$ \textsc{curn}) that the GWB is not present (i.e., a \textsc{curn} signal appears with the same spectrum across the array, but its realizations in different pulsars are spatially uncorrelated);
\item we obtain the observed optimal statistic $D_\mathrm{obs} = D(y_\mathrm{obs})$, and reject $H_0$ if the $p$-value $\int P(D > D_\mathrm{obs}|H_0) \, \mathrm{d}D$ is sufficiently small.
\end{itemize}
As anticipated above, moving beyond the rejection of \textsc{curn} and claiming GWB detection would require additional lines of evidence to conclude that \textsc{hd} is indeed the best explanation for the data.

The ``classic'' optimal statistic is formulated as
\begin{equation}
\label{eq:os}
    D(y) = \frac{\sum_{i \neq j} y^T_i C_i^{-1} \tilde{\Gamma}_{ij} C_j^{-1} y_j}{\sum_{i \neq j} \mathrm{tr} \bigl[ C_i^{-1} \tilde{\Gamma}_{ij} C_j^{-1} \tilde{\Gamma}_{ji} \bigr]}\,,
\end{equation}
where the sum is over all pairs of pulsars in the array;
$y_k$ is the vector of residuals for pulsar $k$;
$C_k$ is the covariance matrix for those residuals, including measurement noise, intrinsic noise, timing-model errors, and common red noise; and $\tilde{\Gamma}_{ij} = \Sigma(t_i,t_j) \times f_\mathrm{HD}(\zeta_{ij})$ is the correlation matrix for GWB-induced residuals in pulsars $i$ and $j$. Here $\Sigma(t_i, t_j)$ is set by the spectral content of the GWB and it is normalized so that in ensemble average $\bigl\langle y^{\phantom{T}}_{i,\mathrm{gw}} \, y^T_{j,\mathrm{gw}} \bigr\rangle = A^2_\mathrm{gw} \tilde{\Gamma}_{ij}$ for a GWB of amplitude $A_\mathrm{gw}$;
while $f_\mathrm{HD}(\zeta_{ij})$ is the Hellings--Downs function of the pulsar-pair angle $\zeta_{ij}$ \cite{hd83}.
See Ref.\ \cite{2014PhRvD..90j4012V} for details about the Gaussian-process formulation of the PTA likelihood.

Under $H_0$, $D(y)$ follows a generalized $\chi^2$ distribution \cite{2023arXiv230501116H} with an expectation value of zero. 
In the presence of the GWB, $D(y)$ follows a non-central generalized $\chi^2$ distribution \cite{2021PhRvD.103f3027R,2022arXiv220807230A}, and its expectation value is $\langle D(y) \rangle = A^2_\mathrm{gw}$. Thus, $D(y)$ may serve as an estimator of $A_\mathrm{gw}$.
Both distributions have been approximated as normal in the optimal-statistic literature:
\begin{equation}
\label{eq:nulldist}
\begin{gathered}
p(D(y) | H_0) \simeq \mathcal{N}(0, \sigma^2_0)\,, \\
\sigma^2_0 = 
\biggl( \sum_{i \neq j} \mathrm{tr} \bigl[ C_i^{-1} \tilde{\Gamma}_{ij} C_j^{-1} \tilde{\Gamma}_{jk} \bigr] \biggr)^{\!\!-1}\,,
\end{gathered}
\end{equation}
and
\begin{equation}
\label{eq:posdist}
p(D(y) | A_\mathrm{gw}) \simeq \mathcal{N}(A^2_\mathrm{gw}, \sigma^2_0)\,.
\end{equation}
For simplicity, we will use these approximations throughout this article, but we caution that their exact distributions \cite{2023arXiv230501116H,2021PhRvD.103f3027R,2022arXiv220807230A} should be used to interpret optimal-statistic results with real data: this is especially important because $p(D(y) | H_0)$ has more substantial tails than the normal distribution, and $p(D(y) | A_\mathrm{gw})$ has variance larger than $\sigma_0^2$ because of GWB-induced correlations between the summands of Eq.\ \eqref{eq:os}.

The $p$-value for $D_\mathrm{obs} \equiv D(y_\mathrm{obs})$ is $1 - \mathrm{CDF}(D_\mathrm{obs}|H_0) \equiv \int P(D > D_\mathrm{obs}|H_0) \, \mathrm{d}D$.
If the normal approximation is taken at face value, it follows from Eq.~\eqref{eq:nulldist} that the combination $\mathrm{SNR} \equiv D(y) / \sigma_0$ can be interpreted as a \emph{signal-to-noise ratio} for null hypothesis testing, so the $p$-value is $\mathrm{erfc}(\mathrm{SNR}/\sqrt{2}) / 2$, where $\mathrm{erfc}$ is the complementary error function (e.g., $p = 0.02, 1.3 \times 10^{-3}, 3.2 \times 10^{-5}, 2.9 \times 10^{-7}$ for $\textrm{SNR} = 2, 3, 4, 5$ respectively).
While this definition of SNR is used broadly in the optimal-statistic literature, it is important to remember that because of the normal approximation an observed SNR of X does not actually imply ``X$\sigma$'' significance.

This optimal-statistic $p$-value refers to the \emph{assumed} population $p(y|\theta_0, H_0)$ of datasets that are generated under the null hypothesis $H_0$ with the assumed pulsar noise parameters $\theta_0$, which enter $D(y)$ through the $C_k$. We will write $D(y;\theta_0)$ to emphasize this dependence.
When analyzing real data, we face the problem that the noise parameters $\theta$ must themselves be estimated from the data.
The simplest approach is setting $\theta$ to their maximum-likelihood or maximum \emph{a posteriori} values $\hat{\theta}(y_\mathrm{obs})$.
Unfortunately, the optimal statistic is very sensitive to pulsar noise assumptions, so fixing them in this way can distort hypothesis-testing conclusions. (It can also lead to biased $A_\mathrm{gw}$ estimates, but that is less of a concern for this article.)
To address this problem,~\citet{2018PhRvD..98d4003V} suggested that we consider the distributions of $D(y_\mathrm{obs};\theta)$ and SNR induced by the Bayesian posterior $p(\theta | y_\mathrm{obs}, H_0)$.
In this approach the SNR gains a notion of Bayesian uncertainty.
hypothesis-testing ``significance'' is quoted as the mean posterior SNR:
\begin{equation}
    \overline{\mathrm{SNR}} = \! \int \! \frac{D(y_\mathrm{obs};\theta)}{\sigma_0(\theta)} \, p(\theta|y_\mathrm{obs}, H_0) \, \mathrm{d}\theta\,.
    \label{eq:meansnr}
\end{equation}
Because the $p$-value is a nonlinear function of the SNR, this average cannot be mapped to the $p$-value of the optimal statistic with respect to any background population. In other words, in performing this marginalization we lose relevant information about the distribution of the optimal statistic, so $\overline{\mathrm{SNR}}$ is not a direct measure of hypothesis-testing significance.

There is however a straightforward way to build a statistically meaningful statistic from the posterior distribution of $D(y_\mathrm{obs};\theta)$: instead of marginalizing the SNR, we marginalize the $p$-value itself.
In addition to making sense intuitively, this procedure admits a principled null-hypothesis-testing interpretation in terms of Bayesian model checking~\cite{rubin1984bayesianly,gelman1996,gelman1996posterior}.
In this framework, the \emph{Bayesian $p$-value} of the observed data is computed over the population of conditional \emph{data replications} generated under $H_0$ with noise parameters $\theta \sim p(\theta|y_\mathrm{obs}, H_0)$:
\begin{multline}
    p_\mathrm{B}(y_\mathrm{obs}) \equiv \int P[D(y^{(\theta)}_\mathrm{rep};\theta) > D(y_\mathrm{obs};\theta)] \\
    \times p(\theta|y_\mathrm{obs},H_0) \, \mathrm{d} \theta\,.
\label{eq:bp}
\end{multline}
where $y^{(\theta)}_\mathrm{rep} \sim p(y^{(\theta)}_\mathrm{rep}|\theta, H_0)$. If we assume that $D(y_\mathrm{obs};\theta)$ is normally distributed, as in Eq.~\eqref{eq:nulldist}, we obtain
\begin{equation}
    p_\mathrm{B}(y_\mathrm{obs}) \simeq \int {\textstyle\frac{1}{2}} \,  \mathrm{erfc}\bigl(\mathrm{SNR}(y_\mathrm{obs};\theta)/\sqrt{2}\bigr) \,
    p(\theta|y_\mathrm{obs},H_0) \, \mathrm{d} \theta \,.
\label{eq:bpapprox}
\end{equation}

The Bayesian $p$-value characterizes how often the null model, \emph{conditioned on the data}, would result in the values of the statistic that we observe---concretely, how often intrinsic pulsar noise with parameter distributions inferred from $y_\mathrm{obs}$ would yield the spatial correlations that we measure.
We propose that $p_B(y_\mathrm{obs})$ should be used to establish the presence of spatial correlations in PTA data by rejecting the null model in hypothesis testing.
We may also map $p_B(y_\mathrm{obs})$ to an effective \emph{Bayesian SNR},
\begin{equation}
\mathrm{BSNR} \equiv \sqrt{2} \, \mathrm{erfc}^{-1}(2 \, p_\mathrm{Bayes})\,,
\label{eq:bsnr}
\end{equation}
thus associating a Gaussian $\sigma$ level with the rejection of $H_0$. Note that the $\mathrm{erfc}$ appears by \emph{definition} in Eq.~\eqref{eq:bsnr}, but is only justified under the normal approximation in Eq.~\eqref{eq:bpapprox}.

The BSNR is skewed toward the lower tail of the posterior $D(y_\mathrm{obs};\theta)$ distribution, because smaller SNRs yield much greater $p$-values, which dominate Eq.~\eqref{eq:bp} and therefore Eq.~\eqref{eq:bsnr}. 
Qualitatively, we are averaging a risk (that \textsc{curn} should yield extreme data), so we pay the most attention to the riskiest pulsar-noise configurations (those that minimize observed correlations).
By contrast, $\overline{\mathrm{SNR}}$ is a direct average of SNR, so it overemphasizes the highest SNRs, which however provide little probability mass to the $p$-value. Thus, BSNR should be quoted instead of $\overline{\mathrm{SNR}}$ as the measure of null-hypothesis-testing significance.

If we do want to use $\overline{\mathrm{SNR}}$, we need to turn to a different scheme.
As for any statistic, the significance of an observed $\overline{\mathrm{SNR}}$ can be obtained \emph{empirically} by sampling its distribution over a relevant population, such as simulations or ``bootstrapped'' data models in which inter-pulsar correlations are masked by randomizing the phases of red-noise Fourier components (\emph{phase shifts}, \cite{tlb+17}) or by randomly assigning pulsar positions when computing the Hellings--Downs function (\emph{sky scrambles}, \cite{cs16,tlb+17}).

These options are correct both formally and substantially, but they answer different questions about the data (because they reference different background populations), so they provide information complementary, but not equivalent, to the BSNR.
Specifically, with simulations we ask how often we would observe a given $\overline{\mathrm{SNR}}$ (or a larger value) over the simulated $H_0$ population, which could have $\theta$ fixed to $\mathrm{argmax}_\theta \, p(\theta|y_\mathrm{obs},H_0)$, or distributed with that posterior. This option is very costly because a new posterior $p(\theta|y_\mathrm{sim},H_0)$ must be obtained for every simulation.
For sky scrambles \cite{cs16,tlb+17}, we randomize pulsar sky positions and ask how often the observed pulsar-pair correlations would happen to conform to the resulting Hellings--Downs patterns (i.e., we explore whether the observed $\overline{\mathrm{SNR}}$ is the product of a lucky sky configuration). For phase shifts \cite{tlb+17}, we ask how often intrinsic red noise with the observed spectral amplitudes, but with random phases, would produce the observed $\overline{\mathrm{SNR}}$.

%

\section{Simulations: SNR distributions and Bayesian p-values}
\label{sec:simulation}
%
\begin{figure}[t]
    \includegraphics[width=\columnwidth]{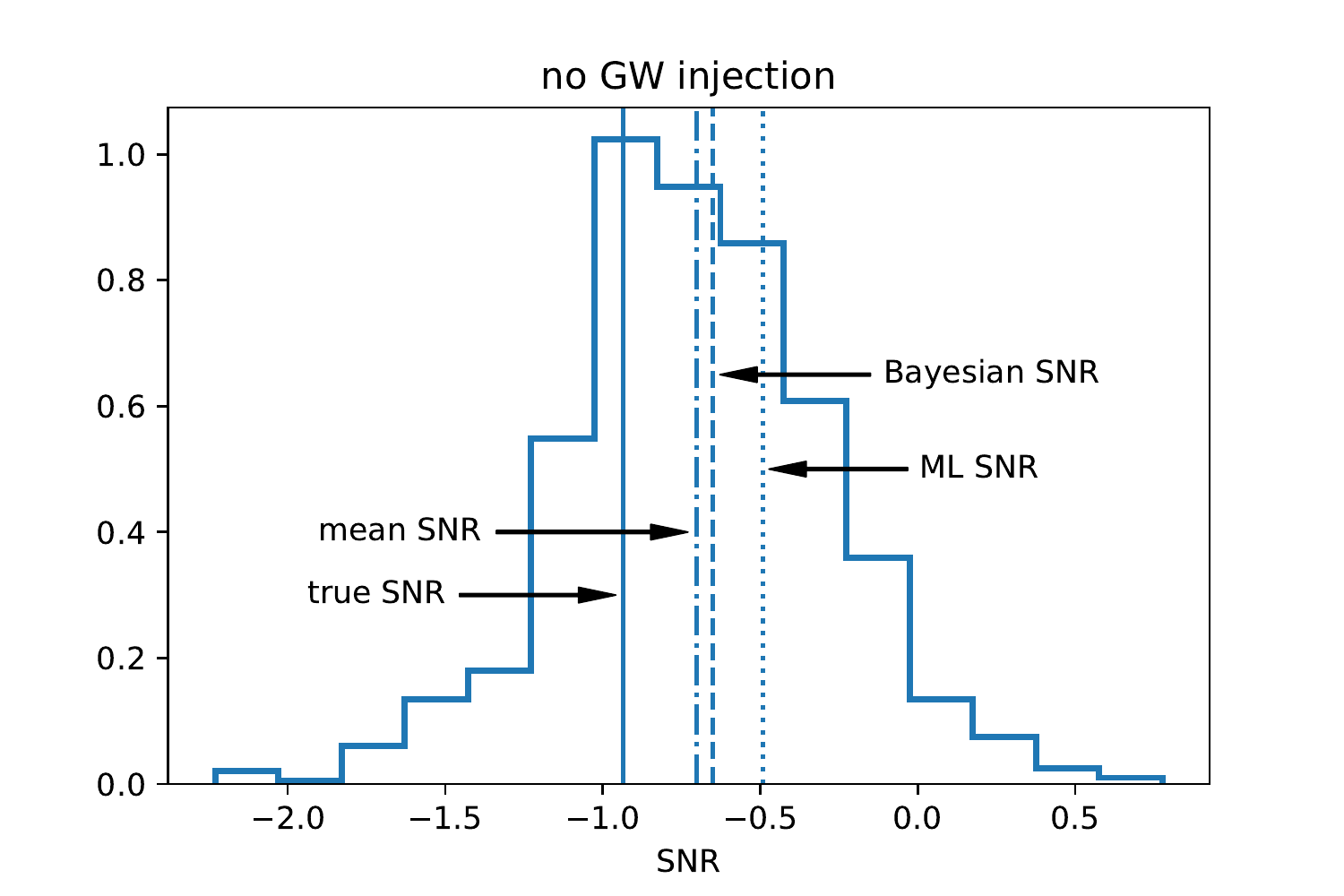}
    \includegraphics[width=\columnwidth]{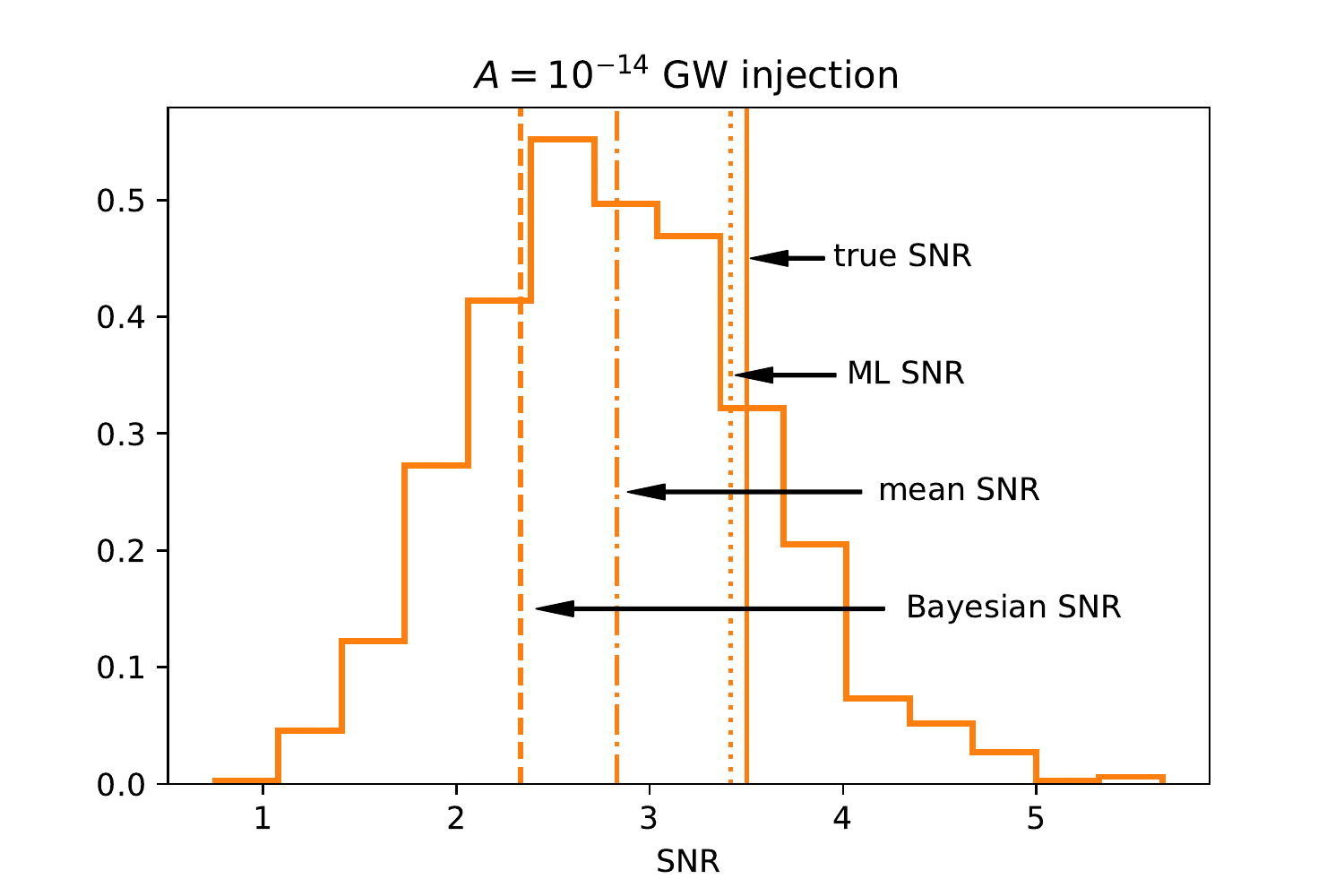}
    \caption{Optimal-statistic SNR posterior distributions for two simulated datasets with no common-process injection (top) and with an $A_\mathrm{gw} = 10^{-14}$ GWB injection (bottom). The no-injection SNR distribution is consistent with the null hypothesis, and the four SNR statistics are close in value. The loud-injection SNR distribution is inconsistent with the null hypothesis, with BSNR $\simeq 3$ and therefore average $p$-value $\simeq 10^{-3}$, significantly lower than implied by the $\overline{\textrm{SNR}}$ and (especially so) by $\mathrm{SNR}(y_\mathrm{obs}; \hat{\theta})$.}
    \label{fig:posteriorsnr}
\end{figure}
In this section we obtain the posterior SNR distribution and compute the BSNR for a set of simulated datasets created to resemble the 12.5-yr NANOGrav data release~\cite{2021ApJS..252....4A}, to get a sense of what we should expect for undetectable and detectable GWBs, and to understand how the BSNR summarizes the distributions. 
All simulations comprise the 47 pulsars in the release, ``observed'' at the same TOAs, but with residuals drawn randomly according to \emph{maximum-a-posteriori} noise hyperparameters $\theta_\mathrm{sim}$ determined from the real dataset. See App.~\ref{app:infsim} for technical details about our simulations and Bayesian inference.

Figure~\ref{fig:posteriorsnr} shows the posterior distribution of $\mathrm{SNR}(y_\mathrm{obs};\theta)$ induced by $p(\theta|y_\mathrm{obs}, \text{\textsc{curn}})$ in two representative simulations: the first (top) with no injection of a GWB or any other common noise, and the second (bottom) with a loud power-law GWB with amplitude $A_\mathrm{gw} = 10^{-14}$ and spectral slope $\gamma = 13/3$ (quoted according to the conventions of~\cite{abb+20}).
The optimal statistic is also built under the assumption of a $\gamma = 13/3$ GWB spectrum.
The vertical bars show $\overline{\textrm{SNR}}$, BSNR, the maximum-likelihood $\mathrm{SNR}(y_\mathrm{obs};\hat{\theta})$, and the \emph{true} $\mathrm{SNR}(y_\mathrm{obs};\theta_\mathrm{sim})$, which of course is not accessible for real data.

The no-injection dataset (top panel) produces a posterior SNR distribution consistent with the null hypothesis, as expected. The four SNR statistics cluster closely. 
By contrast, the loud-injection dataset (bottom panel) produces a posterior SNR distribution that is inconsistent with the null hypothesis, with mean and mode approximately $3.5\sigma$ from zero, close to the true SNR.
The maximum-likelihood SNR is significantly higher at 4.6, and the Bayesian SNR is lower at 2.9. In other words, given the uncertainty in the determination of the noise parameters, we only reach a $p$-value $\simeq 10^{-3}$ ($\simeq 3 \sigma$) rather than the more significant $p$-values incorrectly implied by $\overline{\textrm{SNR}}$ and $\mathrm{SNR}(y_\mathrm{obs}; \hat{\theta})$.

\begin{figure}
 \includegraphics[width=\columnwidth]{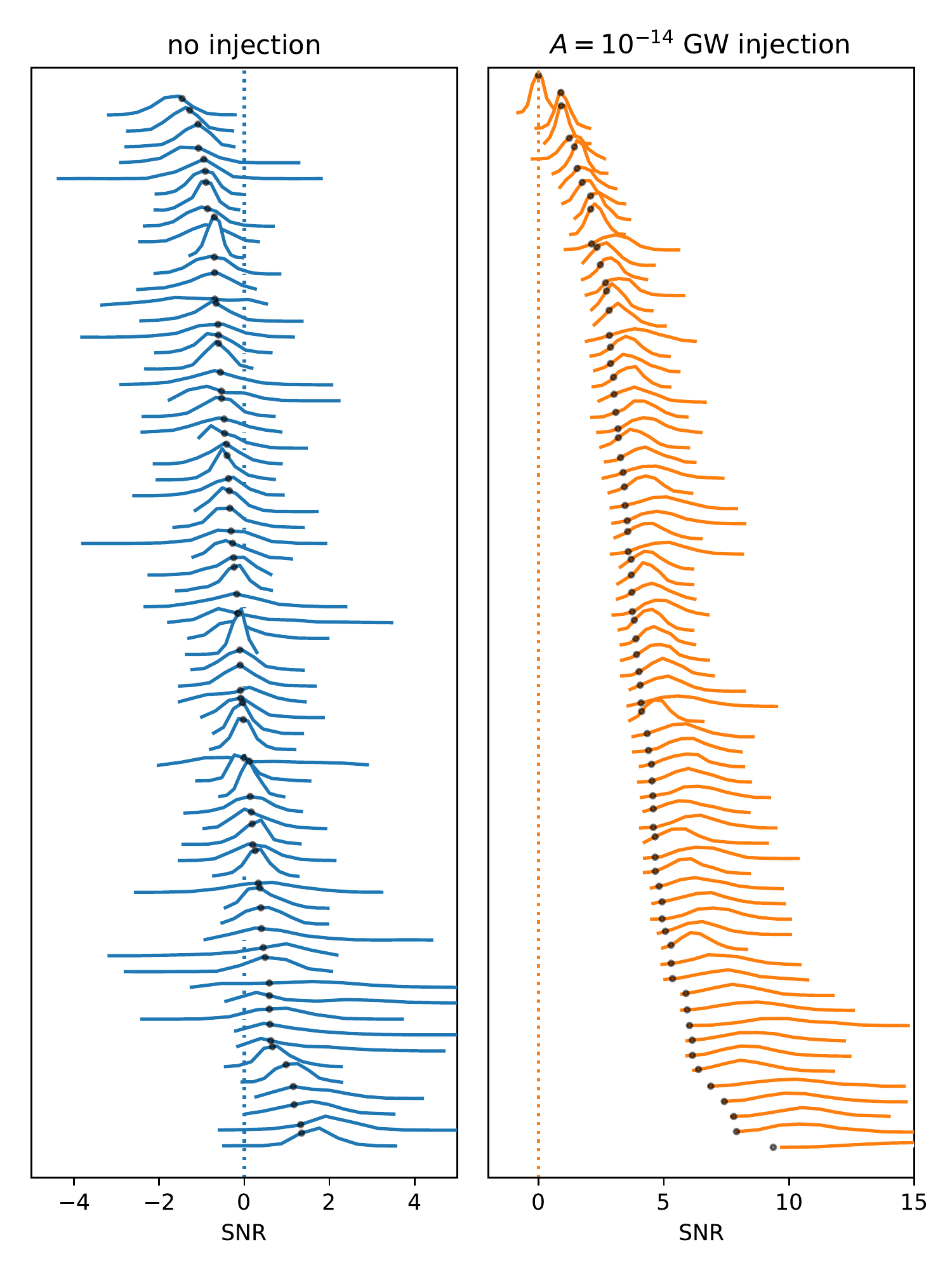}
\caption{Posterior distributions of optimal-statistic SNR for 66 simulated datasets with no GWB (left, blue) and 69 datasets with loud GWB injections (right, orange).
The dots show Bayesian SNRs [Eqs.~\eqref{eq:bp} and~\eqref{eq:bsnr}]. Simulations are sorted according to optimal-statistic SNR.}
\label{fig:hdwaterfall}
\end{figure}
\begin{figure}
\vspace{0pt}\includegraphics[width=\columnwidth]{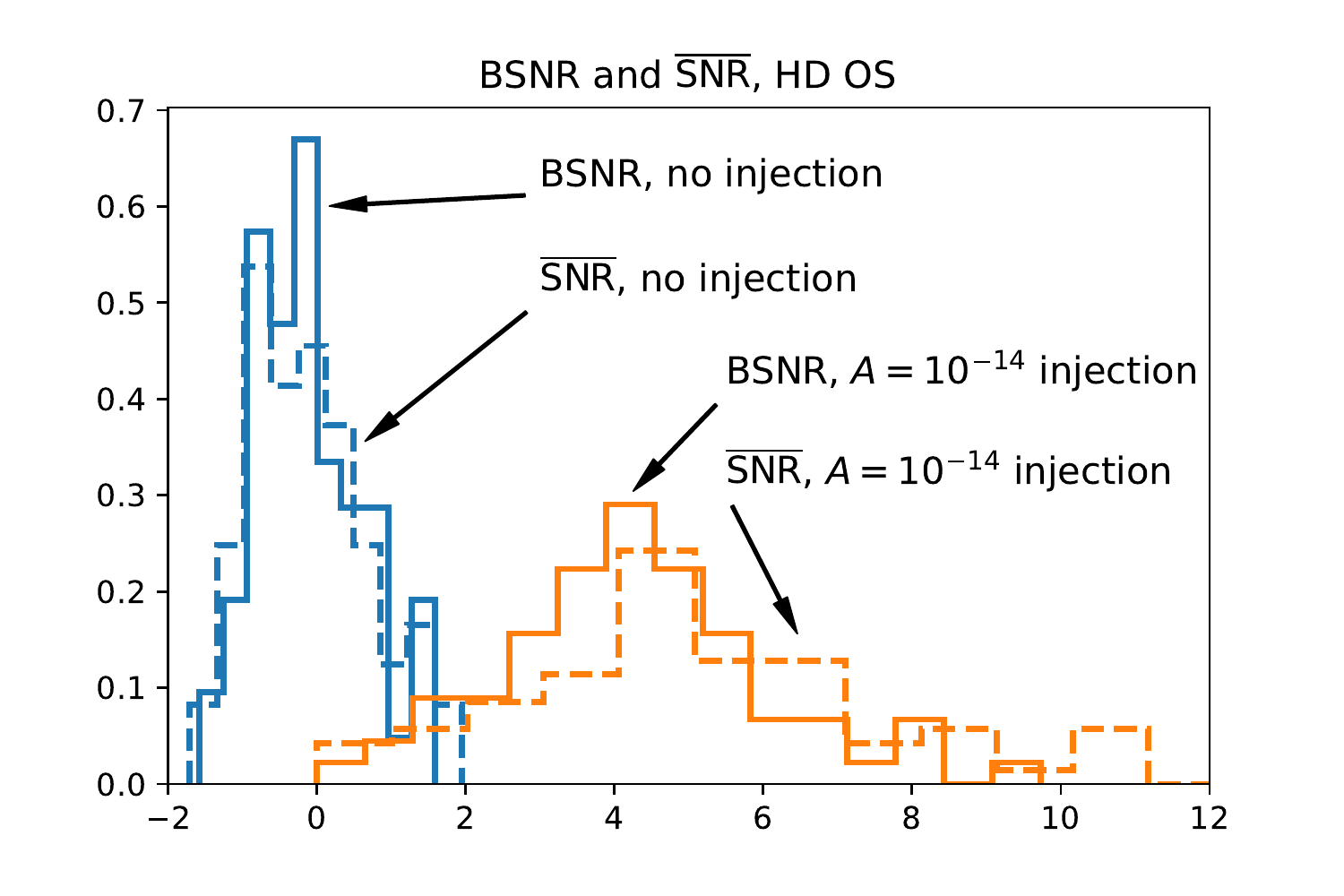}
\caption{
Distributions of optimal-statistic BSNR (solid) and $\overline{\mathrm{SNR}}$ (dashed) over 66 simulated datasets with no GWB (left, blue) and 69 datasets with loud GWB injections (right, orange).}
\label{fig:hdhist}
\end{figure}

Figure~\ref{fig:hdwaterfall} shows posterior optimal-statistic SNR distributions for 66 12.5-yr--like datasets with no GWB or common noise (left, blue), and 69 12.5-yr--like datasets with loud $A_\mathrm{gw} = 10^{-14}$ GWB injections (right, orange).
(The peculiar numbers of datasets resulted from a fraction of simulations failing on our computing cluster because of out-of-memory errors.)
The dots mark BSNR values, which are used to sort the simulations.
Figure~\ref{fig:hdhist} shows histograms of $\overline{\mathrm{SNR}}$ and BSNR across the simulations.

In the no-injection datasets, SNRs cluster around zero as expected,  with a few tails extending to SNRs $\simeq 4$. BSNRs are usually close to $\overline{\mathrm{SNR}}$s.
The blue histograms in Fig.~\ref{fig:hdhist} show the ensemble distributions of $\overline{\mathrm{SNR}}$ and BSNR across simulations, which are concentrated around zero. There is no expectation that BSNRs (or $\overline{\mathrm{SNR}}$s) would be distributed normally \cite{gelman2013}.

In the loud-injection datasets, SNRs are broadly distributed between 0 and 15, with many very convincing detections but also a few false dismissals straddling zero. (These may occur when intrinsic red noise and the GWB pulsar terms, which contribute half of the GWB variance for each pulsar, conspire to obscure the GWB correlations.)
As anticipated in the discussion of Eq.\ \eqref{eq:bp}, the BSNR sits on the lower tail of the SNR distributions (except for the false negatives, in which the tails do not represent extreme $p$-values).

Across simulations (Fig.~\ref{fig:hdhist}, orange histograms on the right), BSNRs cluster around 4, while $\overline{\mathrm{SNR}}$s cluster around 5 with similar tails.
Thus, our admittedly small sample suggests that in approximately half of the realizations of a 12.5-yr-like dataset, the null \textsc{curn} hypothesis could be falsified with $4\mbox{-}\sigma$ Bayesian $p$-value significance, but that the test would be inconclusive ($< 3\mbox{-}\sigma$) in a quarter of realizations.
Incorrectly taking $\overline{\mathrm{SNR}}$s at face value would have overstated significance obviously, if not dramatically, since $\overline{\mathrm{SNR}} > 4$ in 70\% of realizations, and $< 3$ in 17\% of realizations.

Comparing these significance levels with those measured on ``bootstrapped'' background populations is computationally difficult for many of our loud-injection simulations, which have very high significance levels and therefore would require very large backgrounds.
Taking as example one simulation that results in moderate BSNR of 1.9 and $\overline{\mathrm{SNR}}$ of 2.13, we perform 10,000 sky scrambles \cite{cs16,tlb+17} and 10,000 phase shifts \cite{tlb+17}, and find a sky-scramble (phase-shift) significance of 1.9-$\sigma$ (2.4-$\sigma$) for $\overline{\mathrm{SNR}}$. Thus in this case BSNR and sky-scrambled $\overline{\mathrm{SNR}}$ agree. Phase-shifted backgrounds embody less variation, because they effectively fix the amplitudes of red-noise Fourier components and vary only their relative phases, it is reasonable that the phase-shift significance would skew higher.

\section{Interpreting the optimal statistic for alternative correlation functions}
\label{sec:alternative}

GWB searches in PTA data must account for systematic error sources such as long-term oscillations in the time standard~\cite{2020MNRAS.491.5951H} and errors in the Solar System ephemerides~\cite{2020ApJ...893..112V}, both of which are used to refer telescope observations to an inertial frame at rest with respect to the Solar System barycenter.
These errors create timing residuals that are correlated across pulsars, albeit with non-Hellings--Downs geometry: $\langle \delta y_i \delta y^T_j \rangle \propto \tilde{\Gamma}^\mathrm{clk}_{ij} = 1 \times \Sigma(t_i,t_j)$ for clock errors and $\propto \tilde{\Gamma}^\mathrm{ephem}_{ij} = \cos \zeta_{ij} \times \Sigma(t_i,t_j)$ for ephemeris errors [cf.\ the definition of $\zeta_{ij}$ below Eq.~\eqref{eq:os}].
PTA analysts refer to these systematic errors as \emph{monopole} and \emph{dipole}, respectively, with reference to the angular dependence of the inter-pulsar correlations that they embody.
By replacing $\tilde{\Gamma}_{ij}$ with 
$\tilde{\Gamma}^\mathrm{clk}_{ij}$ or $\tilde{\Gamma}^\mathrm{ephem}_{ij}$ in Eqs.~\eqref{eq:os} and~\eqref{eq:nulldist}, we obtain optimal-statistic variants $D^\mathrm{clk}$ and $D^\mathrm{ephem}$ that target monopolar and dipolar correlations, and that have been used to diagnose the presence of correlated systematics in PTA datasets, e.g.,~\cite{abb+20}.
In this section we focus on the monopole optimal statistic, examine its distribution in our simulated datasets, and discuss how Hellings--Downs and monopole signals could be distinguished using the optimal statistic.
\begin{figure}
\includegraphics[width=\columnwidth]{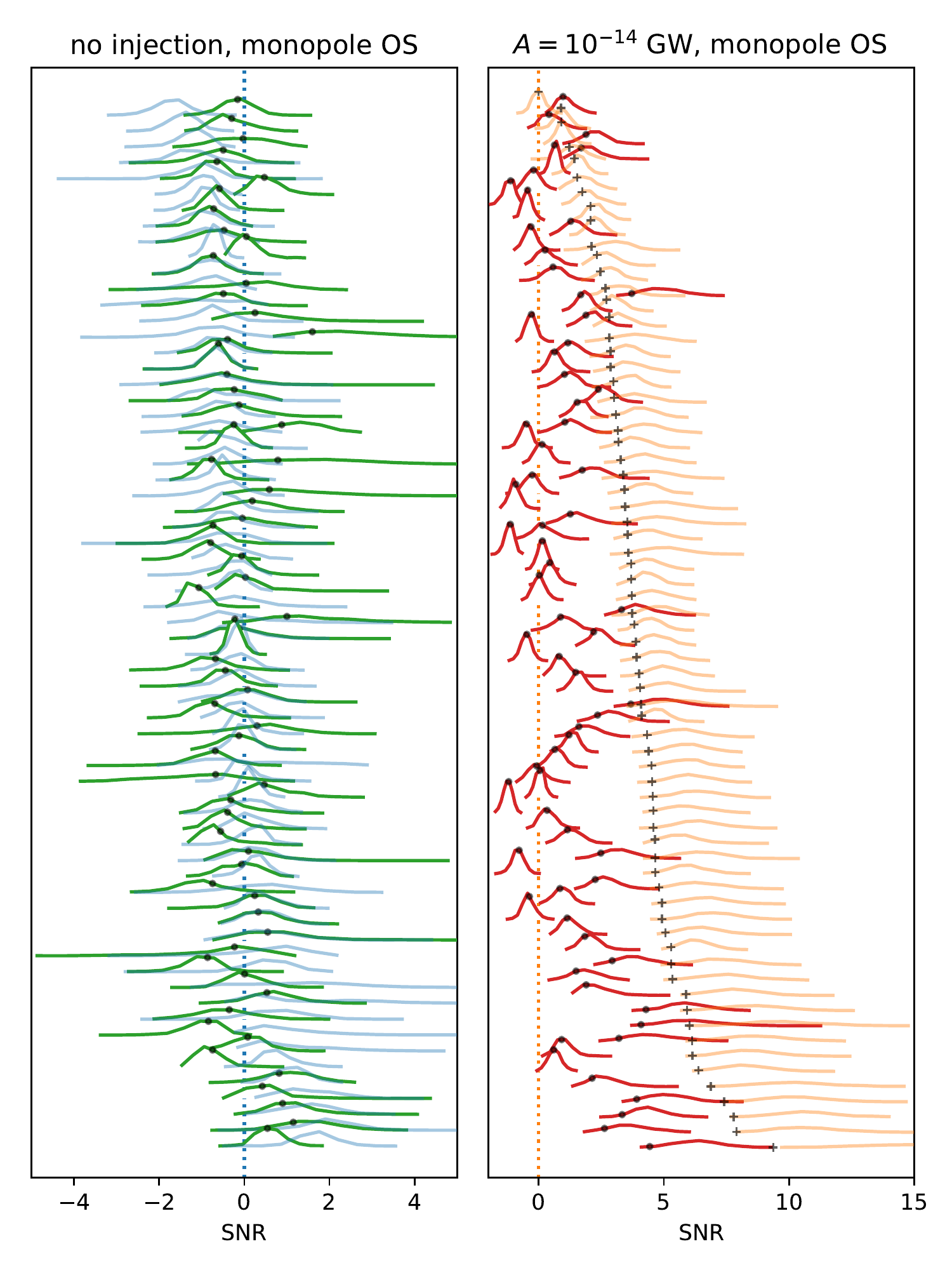}
\caption{Posterior distributions of \emph{monopole} optimal-statistic SNR for 66 simulated datasets with no GWB (left, green) and 69 datasets with loud GWB injections (right, red). Black dots show Bayesian SNRs. The fainter blue and orange curves show the corresponding Hellings--Downs SNR distributions, identical to Fig.~\ref{fig:hdwaterfall}. Simulations are sorted according to the Hellings--Downs optimal-statistic SNR, with plus-shaped markers showing the Hellings--Downs optimal-statistic BSNR.}
\label{fig:monopolewaterfall}
\end{figure}
\begin{figure}
\vspace{0pt}\includegraphics[width=\columnwidth]{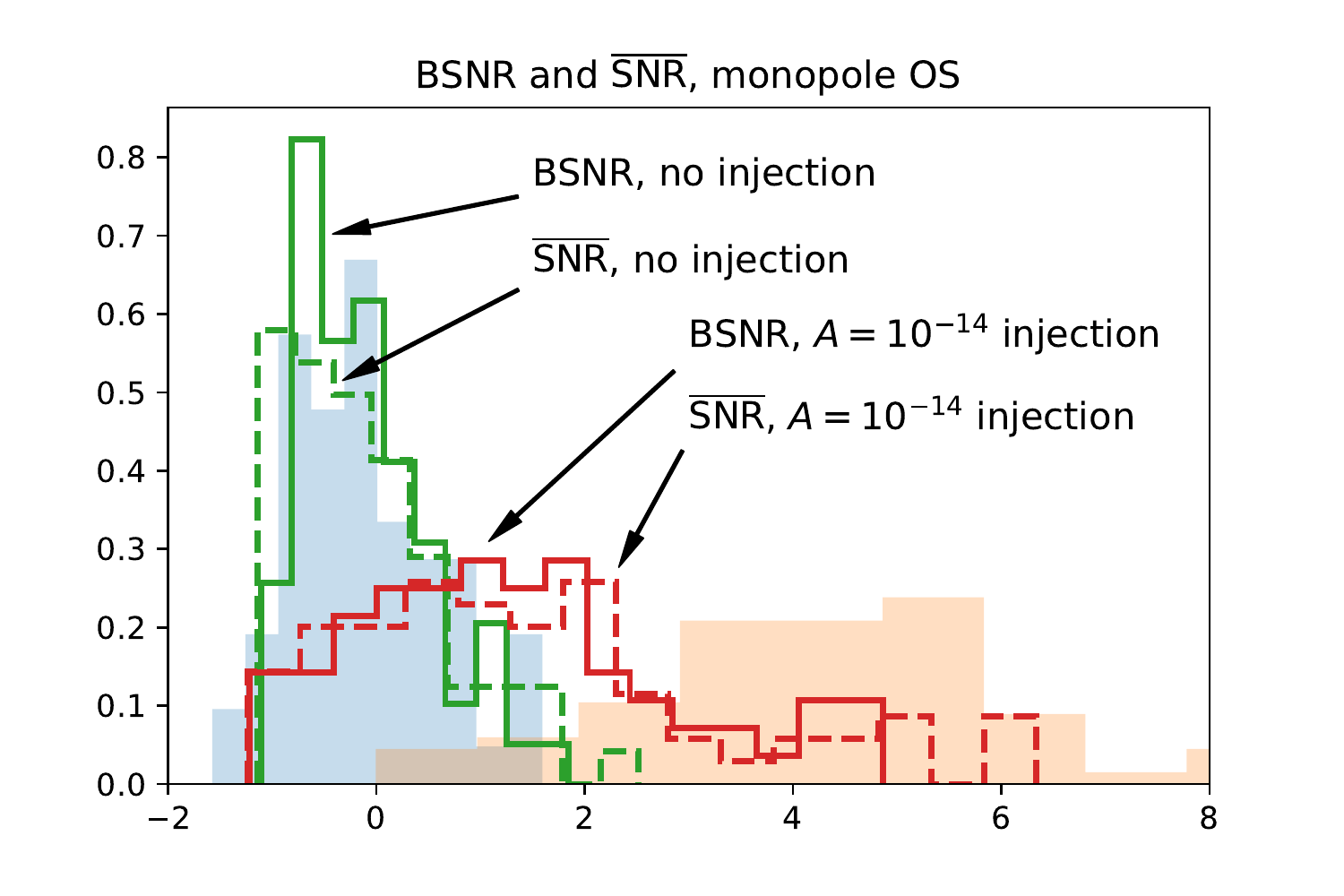}
\caption{Distributions of \emph{monopole} optimal-statistic BSNR (solid) and $\overline{\mathrm{SNR}}$ (dashed) over 66 simulated datasets with no GWB (leftmost, green) and 69 datasets with loud GWB injections (rightmost, red). The fainter blue and orange areas show the corresponding Hellings--Downs BSNR distributions from Fig.~\ref{fig:hdhist}.}
\label{fig:monopolehist}
\end{figure}

Figure~\ref{fig:monopolewaterfall} shows posterior SNR distributions (curves) and Bayesian SNRs (dots) for the monopole optimal statistic in the same no-injection and loud-GWB-injection simulations as Sec.~\ref{sec:simulation}. Datasets are still sorted according to the Hellings--Downs BSNR, and Hellings--Downs SNR distributions are overdrawn more faintly for comparison.
Like the standard optimal statistic, the monopole optimal statistic is built under the assumption of a $\gamma = 13/3$ power law.

In the no-injection datasets, monopole SNRs cluster around zero in a manner similar to the Hellings--Downs optimal statistic; their distributions across simulations are shown in Fig.\ \ref{fig:monopolehist}.
The monopole and Hellings--Downs SNRs are largely uncorrelated, with Pearson $r$ coefficients $\simeq 0.25$ for both $\overline{\mathrm{SNR}}$s and BSNRs.
Overall, the monopole optimal statistic fails (correctly) to reject the null hypothesis.

In the loud-injection datasets (right) monopole SNRs are distributed broadly, although not as much as the Hellings--Downs SNRs, and they extend from $-2$ to 10. Across simulations, monopole BSNRs average to 1.2 and monopole $\overline{\mathrm{SNR}}$s to 1.5; only 4\% of the former and 13\% of the latter are greater than 4 (see Fig.\ \ref{fig:monopolehist}).
Monopole and Hellings--Downs SNRs are significantly correlated, with Pearson $r$ coefficients $\simeq 0.5$ for both $\overline{\mathrm{SNR}}$s and Bayesian SNRs.
Overall, the monopole optimal statistic is much less effective than the Hellings--Downs optimal statistic at rejecting the null hypothesis when a Hellings--Downs-correlated signal is present. Clearly that must be because the statistic encodes the wrong correlation pattern.
Even so, the Hellings--Downs signal \emph{can} excite the monopole statistic, and in some cases (four simulations, or about 6\%) it can even produce $\overline{\mathrm{SNR}}$ and BSNR greater than their Hellings--Downs counterparts. However, none of these four simulations lead to convincing \textsc{curn} rejections.

In other words, a Hellings--Downs signal can still be picked up by the monopole optimal statistic, typically (but not always) with suboptimal SNR. The converse is also true.
Thus, a large monopole SNR does not by itself indicate that systematic residuals with clock-like correlations are present in the data.
More generally, while it is tempting to compare the Hellings--Downs and monopole SNRs to determine which alternative hypothesis is favored by the data, that is not something we can do within null hypothesis testing or its Bayesian extension, in which $p$-values are always computed for the null hypothesis, and therefore carry no quantitative information about the alternatives.
\begin{figure}
    \centering
    \includegraphics[width=0.9\columnwidth]{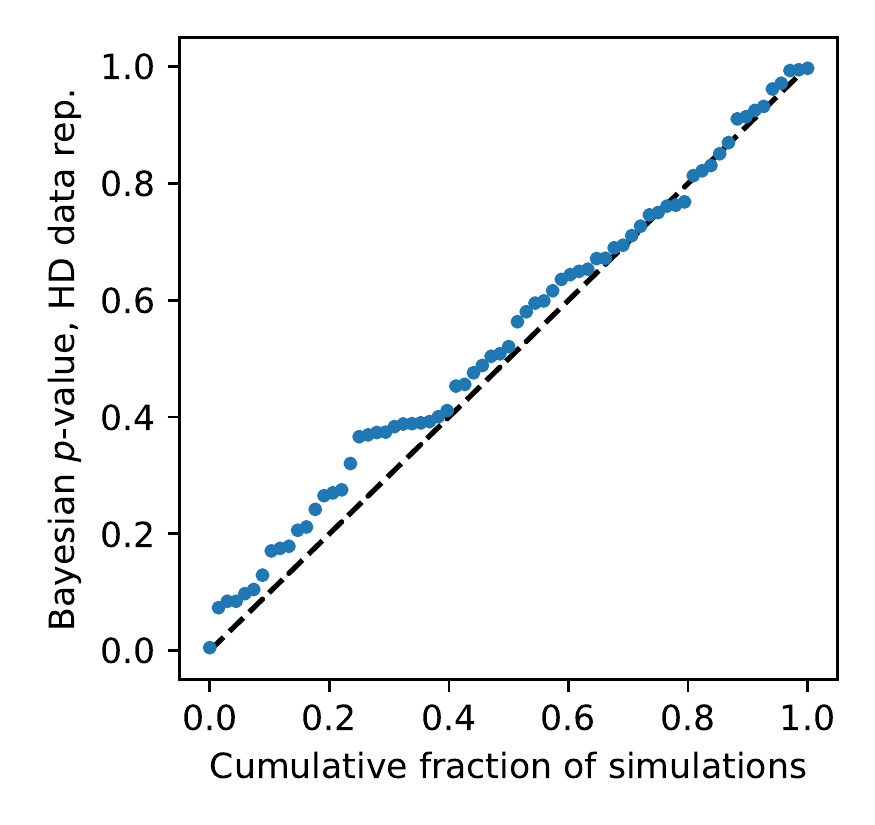}
    \caption{Cumulative distribution of Bayesian $p$-values for $D(y_\mathrm{obs};\theta)$ under the \textsc{hd} hypothesis across 69 loud-injection simulations. As expected, the distribution is approximately uniform. The sampling error of each $p$-value is $\simeq 0.01$.}
    \label{fig:phd}
\end{figure}

In fact, that is the very restriction that we need to address in order to test alternative hypotheses within the optimal-statistic framework.
For instance, having falsified \textsc{curn} because we observed a high Hellings--Downs BSNR, we may now check the \textsc{hd} model by computing the Bayesian $p$-value of $D(y_\mathrm{obs};\theta)$ \emph{under the \textsc{hd} hypothesis}. That is, we consider the distribution of the optimal statistic over \textsc{hd} data replications, modifying Eq.~\eqref{eq:bp} by replacing $p(\theta|y_\mathrm{obs}, \text{\textsc{curn}})$ with $p(\theta|y_\mathrm{obs}, \text{\textsc{hd}})$ and having $y^{(\theta)}_\mathrm{rep} \sim p(y^{(\theta)}_\mathrm{rep}|\theta, \text{\textsc{hd}})$.
The resulting $D(y_\mathrm{rep};\theta)$ is centered on $A_\mathrm{gw}^2$, so we should see that $D(y)$ tracks the estimated GWB amplitude across the posterior $p(A_\mathrm{gw}|y_\mathrm{obs},\text{\textsc{hd}})$.
If \textsc{hd} is the correct hypothesis, we expect to find an unremarkable $p$-value, neither too small nor too close to 1. A very low $p$-value would instead point to mismodeling.

In Fig.~\ref{fig:phd} we perform this check on our loud-injection simulations and find $p$-values that are distributed approximately uniformly between 0 and 1, as expected. A perfect uniform distribution would only obtain in the limit of exact noise-parameter determination, because we would be evaluating the $p$-value of each simulation $y^{(k)}_\mathrm{sim}$ against the true distribution $D(y_\mathrm{sim}; \theta^{(k)}_\mathrm{sim})$.
Because the approximation of Eq.~\eqref{eq:posdist} is inadequate for such strong GWB signals, to build Fig.\ \ref{fig:phd} we evaluate Eq.~\eqref{eq:bp} empirically as
\begin{equation}
\label{eq:empirical}
\frac{1}{NM} \sum_k^N \sum_l^M \Theta\bigl[D(y_\mathrm{obs};\theta^{k}) - D(y^{k,l};\theta^{k})\bigr]\,,
\end{equation}
where $y^{k,l} \sim p(y|\theta^k,\text{\textsc{hd}})$ and $\theta^k \sim p(\theta|y_\mathrm{obs},\text{\textsc{hd}})$.

To check whether we can exclude that the optimal statistic is excited by clock error, we would instead compute the Bayesian $p$-value of $D(y_\mathrm{obs};\theta)$ \emph{under a monopole hypothesis}, using $p(\theta|y_\mathrm{obs}, \text{\textsc{clk}})$ and $p(y^{(\theta)}_\mathrm{rep}|\theta, \text{\textsc{clk}})$ for Eq.~\eqref{eq:bp}.
A small $p$-value would falsify the clock-noise hypothesis, while an unremarkable $p$ would suggest that the model is viable.
Performing this check on our loud-injection simulations yields $p$-values too small too be measured using Eq.\ \eqref{eq:empirical}, but all indeed $\lesssim 0.01$.

\citet{sv2023} propose an extension of the optimal-statistic framework (the multicomponent optimal statistic), in which the pulsar-pair correlations
\begin{equation}
\rho_{ij} = \frac{
y_i^T C_i^{-1} \tilde{\Gamma}_{ij} \, C_j^{-1} y_j
}{
\mathrm{tr} \, (C_i^{-1} \tilde{\Gamma}_{ij} \, C_j^{-1} \tilde{\Gamma}_{ji})}\,,
\end{equation}
are fit to a linear model with components corresponding to different correlation patterns, with errors $\delta \rho_{ij}$ derived under the null hypothesis and assumed to be Gaussian and independent. (In fact, if spatially correlated processes with variance $A_\alpha^2$ are present in the data, they will induce correlations of order $O(A_\alpha^2)$ and $O(A_\alpha^4)$ among the $\rho_{ij}$, which will bias regression unless taken into account iteratively.)

SNRs for each component are defined as the $z$-score of the corresponding linear coefficient.
For a single component, this reproduces the formal SNR of the standard optimal statistic [cf.\ the discussion below Eq.~\eqref{eq:posdist}].
In the general case, the multicomponent optimal statistic attempts to disentangle the cross-sensitivities of the individual optimal-statistic variants, but it can only do so in the context of regression rather than detection.
That is, if we \emph{assume} that a certain set of spatially correlated processes are present in the data, the multicomponent optimal statistic will produce estimates or their relative amplitudes.
When augmented with procedures such as the Akaike information criterion~\cite{Akaike1998}, the multicomponent optimal statistic will also select a best-fitting model among multiple options~\cite{sv2023}.
Neither result can be interpreted easily as rejecting the null hypothesis, or providing quantifiable evidence of an alternative.

\section{The optimal statistic as an approximate likelihood ratio}
\label{sec:plr}

So far we have focused on computing optimal-statistic $p$-values for null hypothesis testing and its Bayesian extension. The $p$-values can falsify the null \textsc{curn} hypothesis, thus confirming the presence of correlations, and
they can also verify that the data are consistent with \textsc{hd}, by failing to falsify that hypothesis.
Moving beyond this strictly falsificationist viewpoint, the optimal statistic is also related (at least approximately) to the ratio of the \textsc{curn} and \textsc{hd} likelihoods~\cite{2013ApJ...769...63E}.
In this section we describe how the likelihood ratio can be used directly to discriminate between the two models, and how it provides a link between the posterior distribution of the SNR and the \textsc{curn}-vs.-\textsc{hd} Bayes factor.
\begin{figure}
    \centering
    \includegraphics[width=\columnwidth]{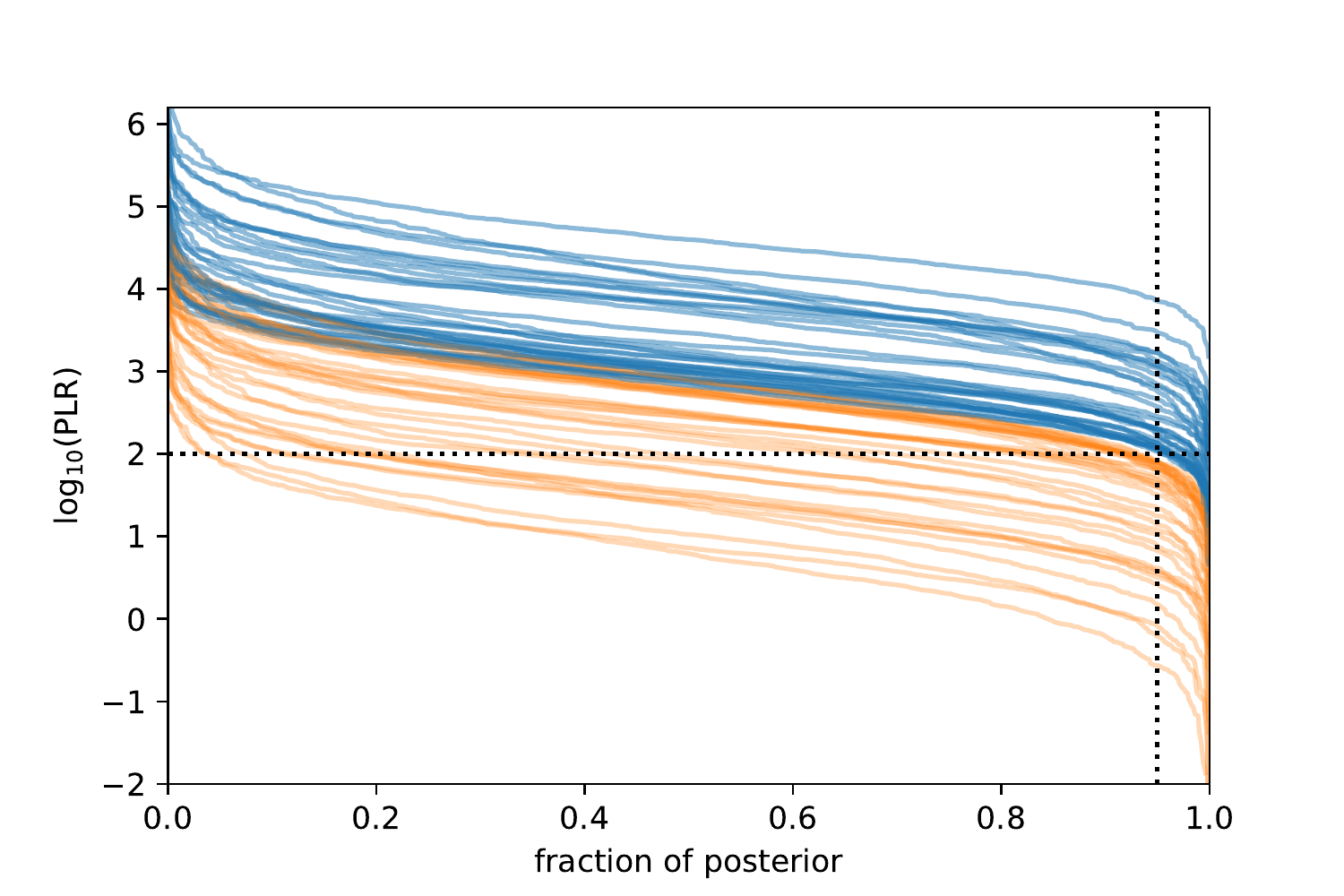}
    \caption{Distribution of the posterior log likelihood ratio (PLLR), as approximated with the optimal statistic using Eq.\ \eqref{eq:approxplr}, in the 69 loud-injection simulations.
    If we set our detection criterion as $\mathrm{PLLR} > 100$ over $95\%$ of the posterior, we conclude that \textsc{hd} is ``detected'' over \textsc{curn} in 48\% of our simulations: those that cross the 95\%-percentile vertical line with values above the $\mathrm{PLLR} = 100$ horizontal line, shown in blue here, while those that do not are shown in orange.}
    \label{fig:plr}
\end{figure}

To see how the optimal statistic is related to the likelihood ratio, we begin with the PTA likelihood in its marginalized form~\cite[see, e.g.,][]{2014PhRvD..90j4012V}:
\begin{multline}
\label{eq:ptalike}
\log p(y|\theta) = -\frac{1}{2}
\mathbf{y}^T \mathsf{K}^{-1} \mathbf{y} - \frac{1}{2} \log |2 \pi \mathsf{K}| \\
\text{with} \; \mathsf{K} = \mathsf{B} + A^2 \tilde{\mathsf{\Gamma}}\,;
\end{multline}
here $\mathbf{y}$ is the concatenation of residuals for all array pulsars; $\mathsf{B}(\theta)$ is a block-diagonal covariance matrix in which each block $B_i$ describes the noise processes that are individual to a pulsar, including measurement noise, intrinsic spin noise, and timing-model errors, but not common red noise; and $A^2 \tilde{\mathsf{\Gamma}}(\theta)$ represents the covariance of the common red-noise process.
For \textsc{curn}, $\tilde{\mathsf{\Gamma}}(\theta)$ is block-diagonal, with blocks given by $\Sigma_{ii}$; for \textsc{hd}, $\tilde{\mathsf{\Gamma}}(\theta)$ has the same diagonal blocks, plus off-diagonal blocks given by $\Sigma_{ij} f_\mathrm{HD}(\zeta_{ij})$.

Expanding $\log p(y|\theta, \text{\textsc{hd}})$ to linear order with respect to the $f_\mathrm{HD}(\zeta_{ij})$ coefficients yields~\cite{2013ApJ...769...63E}
\begin{equation}
\begin{aligned}
\log p(y|\theta, \text{\textsc{hd}}) \simeq & -\frac{1}{2} \sum_{i} y_i^T C_i^{-1} y_i -\frac{1}{2} \sum_i \log |2 \pi C_i| \\
& +\frac{1}{2} A^2 \sum_{i \neq j} y_i^T C_i^{-1} \tilde{\Gamma}_{ij} C_j^{-1} y_j\,,
\end{aligned}
\end{equation}
where $C_i = B_i + A^2 \Sigma_{ii}$. Given that the first two terms in the sum add up to $\log p(y|\theta, \text{\textsc{curn}})$ and that the third term is proportional to the unnormalized optimal statistic [cf.\ Eq.\ \eqref{eq:os}], it follows that 
\begin{equation}
    \frac{
    p(y|\theta, \text{\textsc{hd}})
    }{
    p(y|\theta, \text{\textsc{curn}})
    } \simeq 
    \exp \bigl\{ {\textstyle \frac{1}{2}} A^2 D(y;\theta) / \sigma_0^2(\theta) \bigr\}\,,
    \label{eq:approxplr}
\end{equation}
where of course $A$ is itself a component of $\theta$.

Likelihood ratios are broadly used as detection statistics to discriminate between pairs of models. Indeed, under broad conditions they are Neyman--Pearson-optimal~\cite{1933RSPTA.231..289N}, yielding the lowest rate of false dismissals for a chosen rate of false alarms.
In our context, however, for any given dataset $y$ there is no single value of $D(y; \theta)$ and $\sigma(\theta)$, but rather posterior distributions induced by $p(\theta | y)$.
Likewise, $p(\theta | y)$ induces a posterior distribution of the log likelihood ratio (PLLR).
Dempster~\cite{dempster1997direct,2011AIPC.1305..391S} has suggested that the PLLR can be used directly to make detection statements such as ``under $X$\% of the \textsc{curn} posterior, \textsc{hd} is $Y$ times more likely than \textsc{curn}.''
For instance, one may require $Y = 100$ and $X = 95$\% to claim a detection.

Figure~\ref{fig:plr} shows PLLR distributions for the loud-injection simulations discussed above. Our example criterion is satisfied in 48\% of the simulations.
PLLR statements can be seen as posterior-predictive extensions of classical detection theory. Furthermore, they provide an interesting complement to Bayes factors: instead of ``integrating, then comparing,'' as we do with the Bayesian evidence, with PLLRs we ``compare, then integrate''~\cite{2011AIPC.1305..391S}.
\begin{figure}
    \centering
    \includegraphics[width=0.8\columnwidth]{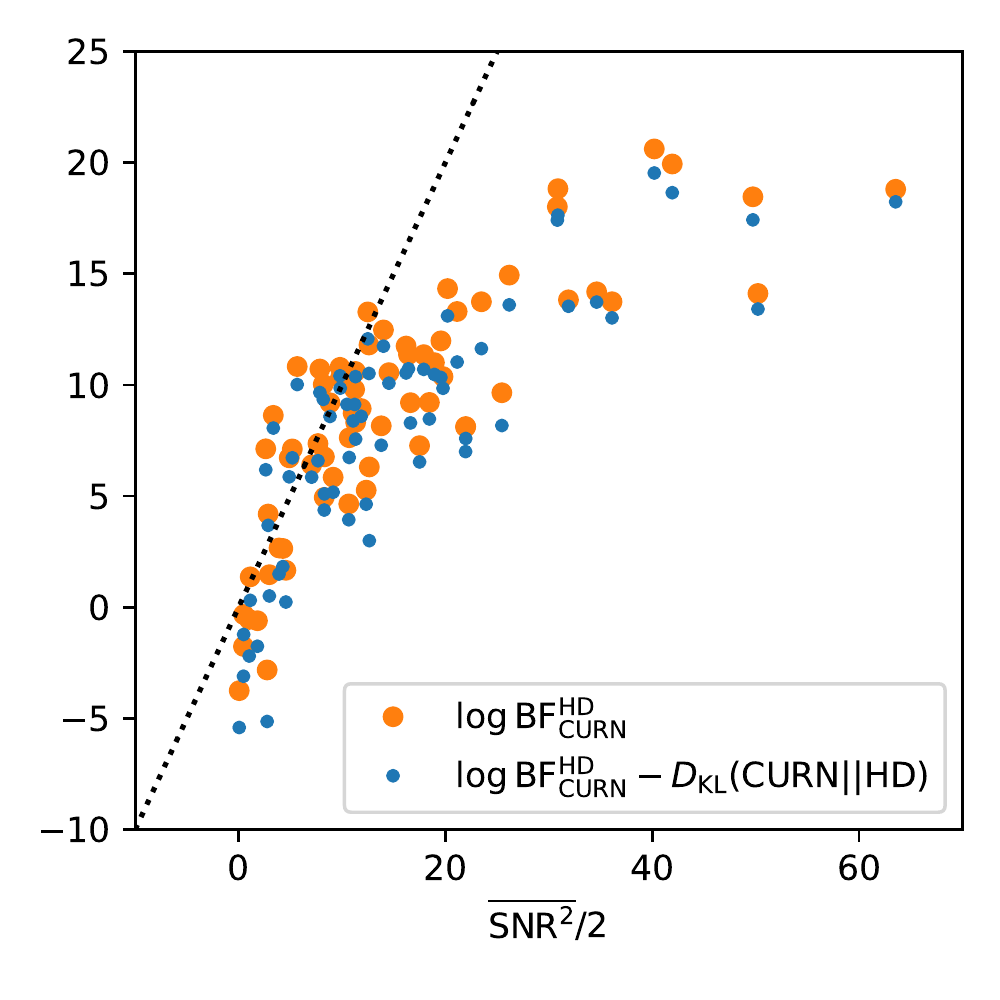}
    \vspace{-12pt}
    \caption{Posterior mean of $\mathrm{SNR}^2/2$ vs log \textsc{hd}-vs.-\textsc{curn} Bayes factors, with and without the correction of Eq.~\eqref{eq:calibrate}, in our loud-injection simulations.
    Bayes factors and Kullback--Leibler divergences are computed using the reweighting scheme of \citet{2023PhRvD.107h4045H}; errors are $\sim 1$, and not shown for clarity.
    Note the large range of $\overline{\mathrm{SNR}^2}$ and Bayes factors, obtained even if all simulations have the same pulsar-noise and GWB parameters.
    Equation~\eqref{eq:calibrate} is realized approximately, with a large vertical spread. Bayes factors appear to saturate toward the high end of the $\overline{\mathrm{SNR}^2}/2$ distribution.}
    \label{fig:snrbf}
\end{figure}

Finally, we demonstrate how the PLLR links the posterior distribution of the optimal-statistic SNR to the \textsc{hd}-vs.-\textsc{curn} Bayes factor. 
 This relationship was examined empirically by Pol and colleagues \cite[Fig.\ 1]{NANOGrav:2020spf}, who find that, for small SNR, $\log \mathrm{BF}^\mathrm{HD}_\mathrm{CURN} \simeq \mathrm{SNR}^2 / 2$.
Because the \textsc{hd} and \textsc{curn} models share the same parameters and priors, the marginalized PLLR yields
\begin{equation}
\label{eq:snrtobf}
\begin{aligned}  
   &\int \log \frac{p(y|\theta,\text{\textsc{hd}})}{p(y|\theta,\text{\textsc{curn}})} p(\theta|y,\text{\textsc{curn}})\, \mathrm{d}\theta \\
   = & \int \log \biggl(
   \frac{p(y|\text{\textsc{hd}})}{p(y|\text{\textsc{curn}})}
   \frac{p(\theta|y,\text{\textsc{hd}})}{p(\theta|y,\text{\textsc{curn}})}
\biggr)
   p(\theta|y,\text{\textsc{curn}})\, \mathrm{d}\theta \\
   = & \, \log \frac{p(y|\text{\textsc{hd}})}{p(y|\text{\textsc{curn}})} \int p(\theta|y,\text{\textsc{HD}})\, \mathrm{d}\theta \, + \\
     & \qquad \qquad \quad \quad \int \log 
   \frac{p(\theta|y,\text{\textsc{hd}})}{p(\theta|y,\text{\textsc{curn}})} \,
   p(\theta|y,\text{\textsc{curn}})\, \mathrm{d}\theta \\
   = & \log \mathrm{BF}^{\text{\textsc{hd}}}_{\text{\textsc{curn}}} - D_\mathrm{KL}(\text{\textsc{curn}}|\!|\text{\textsc{hd}})\,.
\end{aligned}
\end{equation}
In the second line we have used the Bayes theorem plus the fact that
$p(\theta | \text{\textsc{hd}}) = p(\theta | \text{\textsc{curn}})$ to rewrite the likelihood ratio as the Bayes factor times the ratio of posteriors.
The $D_\mathrm{KL}(\text{\textsc{curn}}|\!|\text{\textsc{hd}})$ at the tail end of Eq.~\eqref{eq:snrtobf} is the Kullback--Leibler divergence \cite{kullback1951information} from \textsc{hd} to \textsc{curn}, a
non-negative 
measure of the discrepancy between the two distributions.

Now, if we approximate the PLLR using the optimal statistic [Eq.~\eqref{eq:approxplr}] and replace the amplitude parameter $A^2$ with its optimal-statistic estimator $D(y;\theta)$, we obtain a heuristic relation between the \textsc{hd}-vs.-\textsc{curn} Bayes factor and the optimal-statistic SNR,
\begin{equation}
\label{eq:calibrate}
\log \mathrm{BF}^{\text{\textsc{hd}}}_{\text{\textsc{curn}}} - D_\mathrm{KL}(\text{\textsc{curn}}|\!|\text{\textsc{hd}}) = \overline{\mathrm{SNR}^2}/2\,,
\end{equation}
which \emph{calibrates} the Bayes factor by linking it to a frequentist scheme. For instance, we may say that 3-$\sigma$ optimal statistic corresponds very roughly to $\mathrm{BF} \simeq 90$, while 4-$\sigma$ optimal statistic maps to $\mathrm{BF} \simeq$ 3,000.
Figure~\ref{fig:snrbf} shows that Eq.~\eqref{eq:calibrate} is approximately realized in our loud-injection simulations, in agreement with Ref.\ \cite{NANOGrav:2020spf}.
However, $\overline{\mathrm{SNR}^2}/2$ values consistently overestimate Bayes factors, especially for larger SNRs.
Kullback--Leibler corrections are small.
For this figure, Bayes factors and divergences were computed by reweighting a moderate number of posterior samples~\cite{2023PhRvD.107h4045H}; thus, they are somewhat noisy, but not nearly enough to explain the spread observed in Fig.\ \ref{fig:snrbf}, which must instead originate in the approximations made to obtain Eq.\ \eqref{eq:calibrate}.

Thinking back to the optimal statistic as a detection statistic, the integral over the noise-parameter posterior that defines the Bayesian $p$-value is dominated by the parameters that yield lower SNRs and larger $p$-values, and therefore higher risk; by contrast, Eq.~\eqref{eq:calibrate} contains $\overline{\mathrm{SNR}}^2/2$, emphasizing higher SNRs, and therefore greater confidence.
After all, even if the hypothesis-testing and model-comparison approaches can be related, they answer fundamentally different questions about detection.

\section{Conclusion}
\label{sec:conclusion}

In this article we examine the role of the optimal statistic~\cite{2009PhRvD..79h4030A,2015PhRvD..91d4048C,2018PhRvD..98d4003V}, and especially of its hybrid variant~\cite{2018PhRvD..98d4003V} in establishing the presence of inter-pulsar correlations in pulsar-timing-array data.
The logic is that of null-hypothesis statistical testing: by observing an extreme value of the optimal statistic, we are able to reject a null model that contains a common-spectrum signal but no inter-pulsar correlations. The strength of this conclusion is encapsulated by the $p$-value---the probability that the null model could have produced data that results in an equally extreme statistic.

The fact that we must simultaneously fit for the unknown pulsar noise parameters leads to the hybrid frequentist--Bayesian approach, in which we obtain a posterior distribution for the optimal statistic from the noise-parameter posteriors.
We show that the hybrid optimal statistic can be understood in the framework of Bayesian model checking~\cite{gelman2013bayesian}, in which the $p$-value is evaluated with respect to data replications generated from the null model by drawing model parameters from their posteriors.
This Bayesian $p$-value maps to a new summary statistic, the Bayesian SNR, which should be used to characterize the statistical significance of the observed correlations.
Computed against different posteriors, the Bayesian SNR can also provide principled evidence \emph{for} Hellings--Downs correlations, by failing to reject the \textsc{hd} model, and by rejecting models with alternative correlation patterns such as \textsc{clk}.

By contrast, the posterior SNR of the hybrid SNR (i.e., the $\overline{\textrm{SNR}}$) \emph{cannot} be mapped to a $p$-value for the optimal statistic. Instead, a $p$-value \emph{for the} $\overline{\textrm{SNR}}$ can be established empirically with respect to a population of simulations or of ``bootstrapped'' datasets~\cite{cs16,tlb+17} for which certain model details are altered to effectively erase inter-pulsar correlations.
These are different tests that answer different detection questions with narrower definitions of the null hypothesis, so they should be used in complement to the BSNR.

We also consider the optimal statistic as an approximation to the \textsc{hd}--\textsc{curn} log likelihood ratio.
We suggest that the posterior SNR can be used to formulate detection statements based on Dempster's posterior distribution of the likelihood ratio~\cite{dempster1997direct}.
However these statements may be biased by the approximation of the log likelihood as linear in the Hellings--Downs coefficient, especially so for loud correlated signals at the edge of detection.

Last, we show that the mean \textsc{hd}--\textsc{curn} log likelihood ratio is related to the \textsc{hd}-vs.-\textsc{curn} Bayes factor by way of the Kullback--Leibler divergence between the two posteriors.
Since the ratio is approximately $\overline{\mathrm{SNR}^2}/2$, this relation provides a qualitative calibration of Bayes factors in terms of the hypothesis-testing SNRs that may be expected for similar datasets. The relation also justifies the commonly used heuristic $\log \mathrm{BF} \simeq \mathrm{SNR}^2 / 2$, but our experiments (as displayed in Fig.\ \ref{fig:snrbf}) suggest that the heuristic is realized only very approximately.

\begin{acknowledgments}
This work used the Extreme Science and Engineering Discovery Environment (XSEDE), supported by NSF award ACI-1548562, and specifically the Bridges-2 system at the Pittsburgh Supercomputing Center, supported by NSF award ACI-1928147.
M.V., P.M.M., and K.C., acknowledge support through the NSF Physics Frontiers Center award 1430284 and 2020265.
Part of this research was carried out at the Jet Propulsion Laboratory, California Institute of Technology, under a contract with the National Aeronautics and Space Administration (80NM0018D0004).
Copyright 2023. All rights reserved.
\end{acknowledgments}

\appendix
\section{Inference and simulation}
\label{app:infsim}

Throughout this article, we computed \textsc{curn}, \textsc{hd}, and \textsc{clk} posteriors by evaluating the marginalized PTA likelihood [Eq.~\eqref{eq:ptalike}] with \textsc{Enterprise}~\cite{enterprise}, and sampling it with \textsc{PTMCMCSampler}~\cite{ptmcmc}, following all prescriptions adopted in the NANOGrav 12.5-yr GWB analysis~\cite{abb+20}.
Pulsar noise parameters were set to the maximum \emph{a posteriori}  values obtained in single-pulsar ``noise runs'' on NANOGrav 12.5-yr data~\cite{noisedict}, except for intrinsic--red-noise (log) amplitudes and spectral indices, which were MCMC-sampled alongside $\log_{10} A_\mathrm{curn}$ (or $\log_{10}  A_\mathrm{hd}$ or $\log_{10} A_\mathrm{clk}$, as appropriate).
We used uniform priors of $[-18,-11]$ for the $\log_{10}$ amplitude quantities, and of $[0,7]$ for red-noise spectral indices.  
The spectral index of the common process was fixed to 13/3.
The optimal statistic was evaluated with \textsc{Enterprise} using the matrix components of these likelihoods, drawing intrinsic--red-noise parameters from the appropriate posterior chains, and keeping the other pulsar-noise parameters fixed.

Simulated datasets were obtained under the \textsc{hd} model by fixing \emph{all} noise parameters to 12.5-yr single-pulsar maximum \emph{a posteriori} values, augmented by full-array maximum \emph{a posteriori} values for the intrinsic--red-noise amplitudes and spectral indices. 
To draw random realizations of all noise processes, we decomposed the matrices $B_i$ that appear in Eq.~\eqref{eq:ptalike} as
\begin{equation}
B_i = N_i + F_i \Phi_i F^\dagger_i\,,
\end{equation}
where $N_i$ is an $n^\mathrm{obs}_i \times n^\mathrm{obs}_i$ diagonal matrix (with $n^\mathrm{obs}_i$ the number of measured residuals for pulsar $i$), $\Phi$ is a $n^\mathrm{gp}_i \times n^\mathrm{gp}_i$ diagonal matrix (with $n^\mathrm{gp}_i$ the total number of Gaussian-process basis components for matrix $i$), and $F_i$ is a $n^\mathrm{obs}_i \times n^\mathrm{gp}_i$ the rectangular matrix of Gaussian-process basis vectors.
For each simulated dataset and each pulsar we then obtained
\begin{equation}
    y^\mathrm{noise}_i = \sqrt{N_i} \epsilon + F^\dagger \sqrt{\Phi_i} \zeta\,,
\end{equation}
with $\epsilon$ ($\zeta$) an $n^\mathrm{obs}_i$-dimensional ($n^\mathrm{gp}_i$-dimensional) vector of independent unit Gaussian deviates.
The components of $\Phi_i$ were set to the appropriate power laws for noise processes, and to $10^{-14} \times n_\mathrm{obs} \, \mathrm{s}^2$ for timing-model errors, so that each timing-model parameter could contribute the same variance of $10^{-14} \, \mathrm{s}^2$ (on average) to each residual.
For each simulated dataset the Hellings--Downs process was sampled jointly for all pulsars as
\begin{equation}
    \mathbf{y}^\mathrm{hd} = A \, \mathsf{G} \, \mathsf{L} \, \mathbf{\xi}\,,
\end{equation}
where $\mathsf{G}$ is a block-diagonal matrix in which each $n^\mathrm{obs}_i \times n^\mathrm{hd}$-dimensional block $G_i$ encodes the Hellings--Downs basis vectors for pulsar $i$; $\mathsf{L} \mathsf{L}^\dagger$ is the Cholesky decomposition of the Hellings--Downs covariance matrix $\tilde{\mathsf{\Gamma}}$; and $\mathbf{\xi}$ is an $(n^\mathrm{psr} \times n^\mathrm{hd})$-dimensional vector of independent unit Gaussian deviates.
We set $\log_{10} A_\mathrm{hd}$ to $-18$ for the no-injection datasets and to $-14$ for the loud-injection datasets. Although we ran 100 simulations for each case, only 66 and 69 respectively were completed due to memory limitations on the Bridges-2 computing cluster.

See Refs.~\cite{2014PhRvD..90j4012V,2018ApJ...859...47A,abb+20} for details on the Gaussian-process formulation of the PTA likelihood.

\bibliography{ppc1}

\end{document}